\documentclass[10pt]{article}

\usepackage{geometry}
\usepackage{authblk}
\usepackage[ruled,vlined]{algorithm2e}

\usepackage[english]{babel}
\usepackage[utf8]{inputenc}
\usepackage[colorlinks=true,
			urlcolor=black,
			linkcolor=black,
			citecolor=black]{hyperref}
\usepackage{amsmath, amsthm, amssymb, mathtools}
\usepackage{listings}
\usepackage{graphicx, color, xcolor}
\usepackage{tabularx}
\usepackage{xspace} %list of punctuation marks 
\usepackage{soul} %strike through text

\usepackage[colorinlistoftodos, textwidth = 40mm]{todonotes}

\usepackage{tikz}
\usetikzlibrary{arrows, positioning, automata}	
\tikzstyle{process} = [	rectangle,
						minimum width=1cm, 
						minimum height=0.8cm,
						text centered, 
						draw=black,]
\tikzstyle{arrow} =[thick,->,>=stealth]

\newtheorem{definition}{Definition}[section]

\newcounter{protocol}
\newenvironment{protocol}[1]
  {\par\addvspace{\topsep}
   \noindent
   \tabularx{\linewidth}{@{} X @{}}
    \hline
    \refstepcounter{protocol}\textbf{Protocol \theprotocol} #1 \\
    \hline}
  { \\
    \hline
   \endtabularx
   \par\addvspace{\topsep}}

\newcommand{\G}{\ensuremath{\mathbb{G}}\xspace}
\newcommand{\F}{\ensuremath{\mathbb{F}}}

\newcommand{\kdh}{\mathsf{k_{DH}}}
\newcommand{\ksym}{\mathsf{k}}
\newcommand{\tkdh}{\mathsf{\tilde{k}_{DH}}}
\newcommand{\sk}{\mathsf{sk}}
\newcommand{\pk}{\mathsf{pk}}
\newcommand{\ck}{\mathsf{ck}}
\renewcommand{\pk}{\mathsf{pk}}

\newcommand{\npk}{\mathsf{npk}}
\newcommand{\user}{\mathsf{user}}

\newcommand{\SP}{\mathsf{SP}}
\newcommand{\npkn}[1]{\mathsf{npk}_{#1}}
\newcommand{\nsk}{\mathsf{nsk}}

\newcommand{\stoken}{\mathsf{sc}}
\newcommand{\Note}{\mathsf{N}}

\newcommand{\nftpayload}{\mathsf{payload_{NFT}}}
\newcommand{\sig}{\mathsf{sig}}
\newcommand{\lsig}{\mathsf{sig_{lic}}}
\newcommand{\tsig}{\mathsf{sig_{tx}}}
\newcommand{\attr}{\mathsf{attr}}
\newcommand{\sign}{\mathsf{sign\_single\_key}}

\newcommand{\mintnft}{\mathsf{mint\_nft}}

\newcommand{\nnew}[1]{\mathsf{N^{new}_{#1}}}
\newcommand{\nold}[1]{\mathsf{N^{old}_{#1}}}
\newcommand{\vnew}[1]{\mathsf{v_{#1}}}
\newcommand{\vold}[1]{\mathsf{w_{#1}}}
\newcommand{\change}{\mathsf{change}}
\newcommand{\type}{\mathsf{type}}
\newcommand{\com}{\mathsf{com}}
\newcommand{\enc}{\mathsf{enc}}
\newcommand{\Enc}{\mathsf{Enc}}
\newcommand{\nonce}{\mathsf{nonce}}
\newcommand{\pos}{\mathsf{pos}}
\newcommand{\nullifier}{\mathsf{nullifier}}
\newcommand{\isseen}{\mathsf{is\_seen}}
\newcommand{\pnullifiers}{\mathsf{previous\_nullifiers}}
\newcommand{\lnullifier}{\mathsf{nullifier_{lic}}}
\newcommand{\hb}{H^\mathsf{BLAKE2b}}
\newcommand{\hp}{H^\mathsf{Poseidon}}

\title{Citadel: Self-Sovereign Identities on Dusk Network}
\date{}

\author[1,2]{Xavier Salleras}

\affil[1]{Dusk Network, Amsterdam, Netherlands}
\affil[2]{Universitat Pompeu Fabra, Barcelona, Spain \par
\texttt{xavier@dusk.network}}

\begin{document}

\maketitle

\section*{Abstract}
\noindent 
The amount of sensitive information that service providers handle about their users has become a concerning fact in many use cases, where users have no other option but to trust that those companies will not misuse their personal information. To solve that, Self-Sovereign Identity (SSI) systems have become a hot topic of research in recent years: SSI systems allow users to manage their identities transparently. Recent solutions represent the rights of users to use services as Non-Fungible Tokens (NFTs) stored on Blockchains, and users prove possession of these rights using Zero-Knowledge Proofs (ZKPs). However, even when ZKPs do not leak any information about the rights, the NFTs are stored as public values linked to known accounts, and thus, they can be traced. In this paper, we design a native privacy-preserving NFT model for the Dusk Network Blockchain, and on top of it, we deploy \verb!Citadel!: our novel full-privacy-preserving SSI system, where the rights of the users are privately stored on the Dusk Network Blockchain, and users can prove their ownership in a fully private manner.

\noindent
\\ \textbf{Keywords:} Self-Sovereign Identity; Dusk Network; Zero-Knowledge Proofs; Blockchain; Applied Cryptography.

\pagebreak 
\tableofcontents
\pagebreak 

\section{Introduction}
\label{sec:introduction}

The amount of services available on the Internet is increasing year after year, as well as the concerns about how Service Providers (SPs) handle the sensitive information about their users. In this scenario, SPs are entities that users need to trust, especially regarding the fact that they will behave correctly when managing their personal information. In this regard, Self-Sovereign Identity (SSI) systems \cite{sovrin, sov1, belles2022selfsovereign} have become a research area of interest in the last few years. When it comes to building privacy-preserving authentication methods in services, they play a prominent role: grant users a way to manage their identities in a fully transparent manner. In other words, users of a service implementing an SSI system, are aware of all the sensitive information the SP is requesting, and they can consent to or deny each request.

SSI systems are useful in a wide set of scenarios, like distributing tickets for events, managing users of video or music streaming subscriptions, using car or parking sharing applications, etc. But these examples are not the only kind of contexts where SSI systems are an important privacy feature to implement. If we move our attention to the Internet of Things (IoT) paradigm, we observe a growing density of IoT devices in our cities: medical devices, pollution sensors, traffic lights, IP cameras, etc. In some examples, like an autonomous car sharing its location with an SP or other cars nearby, being able to share only such information but preventing traceability at the very same time, would be an interesting feature to implement. In other words, would be interesting to have SSI systems in these scenarios as well.

\subsection{Related Work} 

Interesting SSI approaches have been introduced recently, like the one detailed in \cite{DBLP:journals/corr/abs-2007-00415}. This paper states a way to deploy an SSI system that grants anonymity to its users at the network level. Other solutions like \cite{abid2022novidchain} introduce a Blockchain-based system for preserving the privacy of users when managing their vaccine certificates.

In most cases, state-of-the-art SSI systems use Zero-Knowledge Proofs (ZKPs) \cite{gmr85} as the backbone of their architecture: cryptographic primitives allowing users to prove knowledge of some information, without leaking anything about it. This is the case of \verb!SANS! \cite{salleras2020sans}, where the authors introduce a private authentication protocol based on these primitives. Using \verb!SANS!, users can prove their rights to access different services, without the SP knowing the identity of the users.

Often, they also rely on Blockchain technologies \cite{blockchain2}, in order to achieve decentralization and immutability, when it comes to buying and granting rights to users. For instance, projects like \textit{iden3}\footnote{https://iden3.io} or \textit{Jolocom}\footnote{https://jolocom.io} build SSI systems where owners of Decentralized Identities (DIDs) are able to manage them in a private manner. At the time of writing this, both solutions rely on the Ethereum Blockchain.

On the other hand, \verb!FORT! \cite{math10040617} is an SSI system that relies on Non-Fungible Tokens (NFTs): assets uniquely identifiable that contain some specific information. What they do, is represent the right acquired by someone as an NFT stored on a Blockchain, and they can prove ownership of this right by means of a ZKP. However, even when it does a great job preserving the privacy of the users of different services, this solution still has some open problems to address: the NFTs, as implemented nowadays, are publicly stored on Blockchains like Ethereum. This means that, even when users can privately prove ownership of such rights, they can still be traced on-chain. As stated in the open problems section of their paper, the authors explain how being able to integrate their solution into Blockchains like Dusk Network would lead to enhanced privacy. Dusk Network is a Blockchain where all the transactions are private by default, and capable of executing smart contracts with built-in privacy features. Being able to integrate a private-by-design NFT model into Dusk, would lead to the possibility of designing and deploying an SSI system on top of it, which would prevent on-chain traceability.

Furthermore, \verb!FORT! presents another problem: the SPs need to be trusted, as the ZKPs sent to them could be reused by them, impersonating like this the users. As such, finding a way to ensure that a license that has been already used cannot be reused in other scenarios, would be a desirable feature.

\subsection{Contributions} 

In this paper, we introduce two main contributions. First, we design a private NFT model to be integrated into Dusk Network. Using such a model, a user buying an NFT will receive a token that only they will be able to read. This approach has full integration with the Dusk Network Blockchain: the changes to the original protocol are minimal and have zero impact on their performance. Plus, our contribution is secure under the same assumptions taken for the original transaction model of Dusk, called Phoenix.

Second, we introduce \verb!Citadel!: an SSI system fully integrated into Dusk that allows users to acquire licenses (a.k.a. rights), and prove their ownership using ZKPs. By means of our novel and private NFT model, the licenses are privately stored in the Blockchain, and thus, we solve the traceability problem that other solutions had. In particular, we provide a system with the following capabilities:

\begin{itemize}
	\item \textbf{Proof of Ownership:} a user of a service is able to prove ownership of a license that allows them to use such a service.
	\item \textbf{Proof of Validity:} our solution introduces the possibility to revoke licenses. Users can prove ownership of a valid license, that has not been revoked.
 	\item \textbf{Unlinkability:} the SP cannot link any activity of their users with other activities done in the network.
 	\item \textbf{Decentralized Nullification:} our system solves the problem regarding the possibility of reusing the proofs, where a malicious SP could impersonate the user after receiving a valid proof: by means of an on-chain and decentralized nullification, like done in the standard stack of Dusk, proofs cannot be reused.
	\item \textbf{Attribute Blinding:} the user is capable of deciding which information they want to leak to the SP, blinding the value and providing only the desired information.
\end{itemize}

Furthermore, our solution is fully integrated into the Dusk stack, where the deployment of the solution will have minimal impact on other parts already implemented. In this same regard, Dusk has some features allowing users to delegate heavy computing tasks to trusted parties, in a secure and private manner. Our solution has been designed taking all these features into account, and thus, heavy computing tasks of our protocol can be delegated as well. This fact is important, as allows for better scalability and faster integration of our protocol into a wider set of scenarios, like web environments, IoT devices with low computing power, etc.

After describing our solution in full detail, we analyze its security, and finally provide benchmarks using our proof-of-concept implementation\footnote{The proof-of-concept implementation can be found in the following repository: https://github.com/dusk-network/citadel}, to demonstrate its deployment feasibility.

\subsection{Roadmap} 

In Section \ref{sec:preliminaries}, we introduce the preliminaries needed to follow up with the whole paper. In Section \ref{sec:buildingblocks}, we introduce in detail Phoenix, the transaction model of Dusk Network, needed to build our solution. In Section \ref{sec:oursolution}, we describe \verb!Citadel! with full details. We conclude and explain the future work in Section \ref{sec:conclusions}.

\section{Preliminaries}
\label{sec:preliminaries}
In this section, we introduce the background needed to understand the whole paper. We first introduce the basics of elliptic curves, commitments, and Merkle trees. Later, we review the specific ZKPs schemes related to our solution, and finally, we describe Blockchain technologies.

\subsection{Elliptic Curves}

One of the main elements required for constructing the ZKP scheme that we will use in this work are elliptic curves. An elliptic curve is defined as follows.

\begin{definition}[Elliptic curve]
    Let $E$ be an algebraic curve defined by the projective solutions of the equation 
    \[Y^2=X^3+aX+b,\] 
    for some $a,b\in\mathbb{F}_{q}$. If $4a^3-27b^2\neq 0$, we call $E$ an \emph{elliptic curve over $\mathbb{F}_{q}$}, and denote this by $E/\mathbb{F}_{q}$.
\end{definition}

We are particularly interested in the so-called pairing-friendly elliptic curves, and we describe them now. Let $E$ be an elliptic curve over a finite field $\mathbb{F}_q$, where $q$ is a prime number. We have the bilinear groups $(\mathbb{G}_1, \mathbb{G}_2, \mathbb{G}_T)$ of prime order $p$, and a pairing 
\[e : \mathbb{G}_1 \times \mathbb{G}_2 \rightarrow \mathbb{G}_T\]
being a bilinear map. As the map $e$ is bilinear, the following relation is satisfied 
\[e(aP,bQ)=e(P,Q)^{ab},\]
for any $P,Q\in E$.

We are particularly interested in two elliptic curves, needed later on to describe the transaction model of Dusk Network. They are the BLS12-381~\cite{zcashBLS} and the Jubjub~\cite{zcashJubJub} elliptic curves. Let $p, q$ be two specific prime numbers of $255$ and $381$ bits, respectively.
The curve BLS12-381 is defined over $\mathbb{F}_q$ by the equation
\[E: Y^2 = X^3 + 4,\]
and has different subgroups $\G_1, \G_2$ such that $\#\G_1 = \#\G_2 = p$. This curve is pairing-friendly, meaning that pairings can be efficiently computed. On the other hand, the Jubjub curve is defined by the equation
\[J : -X^2+Y^2 = 1 + \left(-\frac{10240}{10241}\right)X^2Y^2,\]
over $\mathbb{F}_p$ (it is important to recall that $p$ is the order of a prime subgroup of the BLS12-381). We define a subgroup $\mathbb{J}$ whose order $t$ is a $252$-bit prime. Throughout the document, we will mainly use scalar values from the field $\mathbb{F}_t$, and elements from $\mathbb{J}$.

\subsection{Commitments and Hash Functions}

A \emph{commitment scheme} allows a party to commit to a secret value $v$, to be revealed at a later time. A commitment scheme works like a safe-deposit box, in the following sense. The party that wishes to make a commitment puts the value $v$ inside the box, and locks it. They keep the key, but the box is kept in a public place. The commitment \emph{hides} the value inside, until the owner decides to use the key and open it. At the same time, the commitment \emph{binds} the value, ensuring that the owner cannot change it after committing. We are particularly interested in Non-interactive Commitment Schemes, defined as follows:

\begin{definition}[Non-interactive Commitment]
A non-interactive commitment scheme consists of a tuple of algorithms (Setup, Commit, Open). The Setup algorithm $\ck \leftarrow \mathsf{Setup}(1^\lambda)$ generates a public commitment key $\ck$ given the security parameter $\lambda$. Given the public commitment key $\ck$, the commitment algorithm Commit defines a function $\mathsf{Com}_\ck : \mathcal{M} \times \mathcal{R} \rightarrow \mathcal{C}$ for a message space $\mathcal{M}$, a randomness space $\mathcal{R}$ and a commitment space $\mathcal{C}$. Given a message $m \in \mathcal{M}$, the commitment algorithm samples $r \leftarrow \mathcal{R}$ uniformly at random and computes $\mathsf{Com}_\ck(m;r) \in \mathcal{C}.$ Given $m,r$ and a commitment $c \in \mathcal{C}$, the Open algorithm $\mathsf{Open}_\ck(m;r;c)$ outputs $1/0$ whether or not $c$ is a valid commitment for the pair $m,r$. A non-interactive commitment scheme is perfectly hiding, and computationally binding under the discrete logarithm assumption. 
\end{definition}

In this work, we will use the Pedersen commitment along with the Jubjub elliptic curve. Let $\mathbb{J}$ be a group of order $t$ and set our message and randomness spaces $\mathcal{M}, \mathcal{R} = \F_t$ and our commitment space $\mathcal{C} = \mathbb{J}$. The Setup, Commit, and Open algorithms for Pedersen commitments are defined as follows:
\begin{itemize}
      \item \emph{Setup.} Sample and output the commitment key $\ck= (G,G')\gets\mathbb{J}^2$.
    \item \emph{Commit.} On input a message $m \in \mathcal{M}$, sample randomness $r\gets\F_t$ and output
    \[c = \mathsf{Com}_\ck(m;r)=mG+rG'.\]
    \item \emph{Open.} Reveal $m,r$. With these values, anyone can recompute the commitment and check whether it matches the commitment previously provided.
\end{itemize}

The Pedersen commitment scheme is perfectly hiding, and computationally binding under the discrete logarithm assumption.

On the other hand, we will make use of hash functions, which we define as follows.

\begin{definition}[Hash Functions] A \emph{cryptographic hash function} is a function $H:\{0,1\}^*\rightarrow\{0,1\}^\ell$ that is collision-resistant, that is, it is hard to find $x,x'\in\{0,1\}^*$ such that $x\neq x'$ but $H(x)=H(x')$. 
\end{definition}

Throughout the document, we will be using two specific hash functions: first, BLAKE2b \cite{blake2}, a lightweight and efficient hash function. Second, Poseidon \cite{cryptoeprint:2019:458}, whose main feature is being SNARK-friendly: it is cheap in terms of computing resources when computed into a specific ZKP scheme called zk-SNARK.

\subsection{Merkle Trees}

\emph{Merkle trees}~\cite{merkle1987digital} are data structures containing at every node the hash of its children nodes. Considering a $k$-ary tree of $h$ levels, the single node at level $0$ is called the \emph{root} of the tree, and the $k^{h}$ nodes at level $h$ are called the \emph{leaves}. Given a node placed in the level $i$, the $k$ nodes in the level $i+1$ that are adjacent to it are called its \emph{children}. Plus, a node is the other's \emph{sibling} if they all are children of the same node.

The tree is partially updated every time a new value is written (or modified) into a leaf, always resulting in a new root of the tree. Furthermore, given a root $r$, it is easy to prove that a value $x$ is in a leaf of a tree with root $r$. The proof works as follows: 
\begin{itemize}
  \item \emph{Prove.} For $i=h,\dots,1$, let $x_i$ be the node that is in level $i$ and is in the unique path from $x$ to the root. Let $y_{i, 1},\dots, y_{i, k-1}$ be the $k-1$ siblings of $x_i$. Output
  \[(x, (y_{1,1}, \dots, y_{1,k-1}), \dots, (y_{h,1},\dots,y_{h,k-1})).\]
  \item \emph{Verify}. Parse input as $(x_h, (y_{1,1}, \dots, y_{1,k-1}), \dots, (y_{h,1},\dots,y_{h,k-1}))$, where $x_h$ is the purported value and $y_{i,1},\dots,y_{i,k-1}$ are the purported siblings at level $i$. For $i=h-1,\dots, 0$, compute\footnote{Additionaly, the prover also has to send $\lceil\log_2k\rceil$ bits for each level, specifying the position of $x_i$ with respect to its siblings, so that the verifier knows in which order to arrange the inputs of the hash.}
  \[x_i = H\left( x_{i+1}, y_{i+1,1},\dots,y_{i+1,k-1} \right).\] 
  If $x_0$ equals the root $r$ of the set we are proving membership of, the proof is verified.
\end{itemize}

We can prove membership in a set of size $k^h$ by sending $kh$ values, so we can state that the communication complexity for proving the membership is $O(kh)$. If the hash function is collision-resistant, the proof is sound.

\subsection{Digital Signatures}
\label{sec:signatures}

Digital signatures are one of the most important pieces needed to build our solution. In particular, we are interested in the Schnorr signature scheme, which we describe now. Let $G, G'\gets \mathbb{J}$. Then, we have the following algorithms:

\begin{itemize}
    \item \emph{Setup.} Sample a secret key $\sk \gets \F_t$ and compute a public key $\pk = \sk G$. 
    \item \emph{Sign.} To sign a message $m$ using $\sk$, sample $r \gets \F_t$ and compute $R = rG$. Compute the challenge $c = H(m, R)$, and set 
    \[u = r - c \sk.\]
    Set the signature $\sig = (R, u)$. 
    \item \emph{Verify.} To verify a signature $\sig = (R, u)$ of a message $m$ using $\pk$, we first compute $c = H(m, R)$ and check whether the following equality holds:
    \[\begin{aligned}
        & R \stackrel{?}{=} uG + c \pk, \\
    \end{aligned}\]
    If it equals, accept the signature, reject otherwise.
\end{itemize}

This scheme is existentially unforgeable under chosen-message attacks under the discrete logarithm assumption, in the random oracle model~\cite[Section 12.5.1]{katz2020introduction}.

We are also interested in a double-key version of this signature scheme, that will be used to delegate some computations later in the protocol. We have the following algorithms:

\begin{itemize}
    \item \emph{Setup.} Sample a secret key $\sk \gets \F_t$ and compute a public keypair $(\pk, \pk') = (\sk G, \sk G')$. 
    \item \emph{Sign.} To sign a message $m$ using $\sk$, sample $r \gets \F_t$ and compute $(R, R') = (rG, rG')$. Compute the challenge $c = H(m, R, R')$, and also 
    \[u = r - c \sk.\]
    Finally, set the signature $\sig = (R, R', u)$. 
    \item \emph{Verify.} To verify a signature $\sig = (R, R', u)$ of a message $m$ using $(\pk, \pk')$, we first compute $c = H(m, R, R')$ and check whether the following equalities hold:
    \[\begin{aligned}
        & R \stackrel{?}{=} uG + c \pk, \\
        & R' \stackrel{?}{=} uG' + c \pk'. \\
    \end{aligned}\]
    If they are equal, accept the signature, reject otherwise.
\end{itemize}

\subsection{Zero-Knowledge Proofs}
\label{subsec:zkp}

A Zero-Knowledge Proof (ZKP) \cite{gmr85} is a cryptographic primitive allowing a prover $\mathcal{P}$ to convince a verifier $\mathcal{V}$ that a public statement is true, without leaking any secret information. 

Given a statement $u$, and a witness $w$ being some secret information only known by $\mathcal{P}$, $\mathcal{P}$ wants to convince $\mathcal{V}$ that they know $w$. Both $u$ and $w$ are related by a set of operations defined by a \textit{circuit}, a graph composed of different wires and gates, which leads to a set of equations involving the inputs and the outputs of these gates. Each of these equations is called a \textit{constraint}. $\mathcal{P}$ can execute a proving algorithm using $u$ as the set of public inputs, and $w$ as the private inputs. This execution outputs a set of elements, which we call the proof $\pi$. $\mathcal{P}$ sends $\pi$ to $\mathcal{V}$, who will use a verifying algorithm to verify that $u$ is true, for a given $w$ only known by $\mathcal{P}$. In essence, ZKPs must satisfy 3 properties:

\begin{itemize}
 \item \textbf{Completeness:} If the statement is true, $\mathcal{P}$ must be able to convince $\mathcal{V}$.
 \item \textbf{Soundness:} If the statement is false, $\mathcal{P}$ must not be able to convince $\mathcal{V}$ that the statement is true.
 \item \textbf{Zero-knowledge:} $\mathcal{V}$ must not learn any information from the proof beyond the fact that the statement is true.
\end{itemize}

First ZKP schemes used to achieve the aforesaid properties by exchanging several messages between $\mathcal{P}$ and $\mathcal{V}$. However, non-interactive ZKPs \cite{Blum:1988:NZA:62212.62222} arose, providing an extra feature, where $\mathcal{P}$ could prove statements to $\mathcal{V}$ by sending them a single message, instead of several interactions. 

Even so, computing and verifying ZKPs used to require high computing resources, and this made them impractical in real applications. More recently, Zero-Knowledge Succinct and Non-interactive ARguments of Knowledge (zk-SNARKs) \cite{cryptoeprint:2013:879} appeared: ZKPs that can be computed and verified in a more efficient way, compared to previous solutions, making them suitable for real applications, like privacy-preserving cryptocurrencies \cite{zcash}. 

\subsection{zk-SNARKs}

zk-SNARKs \cite{184425} are the most used ZKPs, because they are short and succinct: the proofs can be verified in a few milliseconds. However, they require a trusted setup where some public parameters are generated. These parameters, called the Common Reference String (CRS) are used by $\mathcal{P}$ and $\mathcal{V}$ to generate and verify proofs. To generate the CRS, a secret randomness $\tau$ is used, and such randomness should be destroyed afterward. If an attacker gets $\tau$, the soundness property of the scheme breaks: the attacker would be able to compute false proofs that anyone could verify as if they were correct. As such, the CRS is commonly computed using a secure Multi-Party Computation (MPC) protocol \cite{cryptoeprint:2017:1050}, where $\tau$ can only be leaked if all the participants are malicious. The computing complexity of generating a setup, computing proofs, and verifying them, depends on the number of operations that we do in the circuit, which is also the number of gates $n$. 

\subsection{Blockchain}

A Blockchain \cite{blockchain} is a unique and immutable data structure called ledger, and shared by a set of nodes. Cryptocurrencies like Bitcoin \cite{blockchain2} use such technology, and populate the ledger with transactions exchanging money between parties. These transactions are cryptographically validated by the nodes of the network, to be sure that each user spends what belongs to them. This process is a consensus agreed upon among all the users of the network (e.g. Proof of Work \cite{gervais2016security}, Proof of Stake \cite{bentov2014proof}, etc.). 

Beyond the feature of exchanging money, Blockchains like Ethereum \cite{ethereum} grant the possibility of executing decentralized applications (DApps) on-chain. DApps are possible thanks to smart contracts \cite{mavridou2018designing}, programs that can be executed on-chain thanks to the Ethereum Virtual Machine (EVM) \cite{hildenbrandt2018kevm}. Such contracts and the EVM allow, for instance, to execute some action (like issuing a payment) upon fulfilling some conditions.

Furthermore, other Blockchains like Dusk Network \cite{dusk} also provide virtual machines to execute smart contracts. In this particular case, Dusk has the \verb!Rusk! virtual machine, which like the EVM can execute smart contracts, but with the difference that all the transactions handled by \verb!Rusk! are private by default, thanks to ZKPs.

From a more technical perspective, it is worth mentioning that, in order to prevent saturation of the network, users are required to pay \textit{gas} in order to execute transactions. This is the amount of Duskies (Dusk's coin) per amount of bytes needed to execute a transaction. Depending on how busy the Dusk Network is, the price of the gas increases or decreases. Like this, performing a Denial-of-Service (DoS) attack becomes so expensive that is infeasible \cite{Chen2017AnAG}.

\section{Phoenix Transaction Model}
\label{sec:buildingblocks}

In this section, we introduce the details about Phoenix\footnote{https://github.com/dusk-network/phoenix-core}, the transaction model used by Dusk Network. 

\subsection{Overview of Phoenix}

Dusk Network is an open-source public blockchain with a UTXO-based architecture that allows the execution of obfuscated transactions and confidential smart contracts. In Phoenix, UTXOs are called {\textit{notes}}, and the network keeps track of all these notes by storing their hashes in the leaves of a \textit{Merkle tree of notes}. In other words, when a transaction is validated, the network includes the hashes of the new notes to the leaves of this tree. 

All transactions include a ZKP called \textsf{tx\_proof} that proves that the transaction has been performed following the network rules. Essentially, what this ZKP does is the following: first, it nullifies a note that the user is willing to spend. Second, proves that the user knows the value of a new note to mint, that will be sent to the receiver. Finally, proves that the amount of Dusk coins nullified is equal to the amount of coins created.

Greater details about the parameters included in the transaction have been skipped here, for the sake of completeness. Nevertheless, the next subsections explain how the protocol manages the notes nullification and minting, by introducing first the different keys the users have to handle.

\subsection{Protocol Keys}
\label{sec:protocol-keys}

In Phoenix, we have different kinds of keys. First, we have the static keys that belong to each user of the network, and we introduce them here as follows. Let $G, G'\gets\mathbb{J}$ be two JubJub points acting as our generators. We denote by $\hp$ and $\hb$ the Poseidon and BLAKE2b hash functions, respectively. Each user computes the following keys:

\begin{itemize}
	\item \textbf{Secret key:} $\sk = (a,b)$, where $a,b\gets \F_t$.
	\item \textbf{Public key:} $\pk = (A,B)$, where $A = a G$ and $B = b G$.
\end{itemize}

As noticed, Phoenix uses two-element keys, which allows users of the network to delegate the process of scanning for the notes addressed to them.

On the other hand, each note is associated with a unique one-time key pair (an approach introduced in~\cite{van2013cryptonote}), instead of using the static public key of the receiver, which hinders traceability. 

The computation of these keys is based on the Diffie--Hellman key exchange protocol \cite{diffie1976new}. The note public key of a note sent to a receiver with public key pair $\pk = (A,B)$, and its associated note secret key, are computed as follows:

\begin{itemize}
	\item \textbf{Note public key:} a sender willing to send money to a receiver whose public key $\pk = (A,B)$ is known by them in advance, must first compute a note public key $\npk$ following next steps.

	\begin{enumerate}
		\item Sample $r$ uniformly at random from $\F_t$.
		\item Compute a symmetric Diffie--Hellman key $\kdh = rA$.
		\item Compute a one-time public key $\npk = \hb(\kdh)G + B$.
		\item Compute $R = rG$.
	\end{enumerate}
	
	The sender of a note will attach to it the note public key $\npk$ and the partial Diffie--Hellman $R$ used to create $\npk$. Given a pair $(\npk, R$), the receiver can identify whether the note was sent to them by recomputing $\tkdh = aR$ (using their secret $a$), and checking the equation 

	\[\npk \stackrel{?}{=} \hb(\tkdh) G + B.\]
	
	\item \textbf{Note secret key:} the receiver can compute the note secret key $\nsk = \hb(\kdh) + b$, to be used when willing to spend that note. This key can only be computed by the receiver of the note since they are the only ones holding the whole secret key $\sk=(a,b)$, and $\sk$ cannot be recovered from public information. This is due to the discrete logarithm assumption in $\mathbb{J}$.
\end{itemize}

\subsection{Protocol Details}
\label{sec:transaction-model}

A note is defined as the following set of elements:

$$\Note = \{\type, \com, \pos, \nonce, \enc, \npk, R\}.$$

where $\type$ indicates the type of the note, either transparent or obfuscated; $\com$ is a commitment to the value of the note; $\pos$ is the position of the note in the Merkle tree of notes; $\nonce$ is an initialization vector needed for the encryption scheme; $\enc$ is an encryption of the opening of $\com$ that can be decrypted using the receiver's view key; $\npk$ is the note's public key, whose associated private key $\nsk$ can only be computed by the receiver of the note; and $R$ is a point in the Jubjub subgroup $\mathbb{J}$ that allows the receiver to compute $\nsk$ and also identify that they are the receiver of the transaction.

We describe Phoenix in the general scenario in which a sender wishes to send different amounts of money $\vnew{1},\dots,\vnew{n}$ to different receivers with public keys $\mathsf{pk_1},\dots,\mathsf{pk_n}$. We assume the sender owns a set of notes $\{\nold{1},\dots,\nold{m}\}$ each with an associated amount $\vold{i}$ such that \[\sum_{i = 1}^m \vold{i} \geq \sum_{i = 1}^n \vnew{i},\] i.e. the sender has enough funds.

When creating the transaction to transfer the funds, the sender will have to nullify the set of old notes being spent and mint a new set of notes $\{\nnew{1},\dots,\nnew{n}\}$ with the corresponding values $\vnew{i}$, and assigned to the corresponding receivers. 
The most common case is $\mathsf{n} = 2$, where a sender generates a note $\nnew{1}$ with value $\vnew{}$ for a receiver, and a second note $\nnew{2}$ for themselves with value  
$\change = \vnew{} - \sum_{i=1}^m\vold{i}$. 

To mint a new note for a receiver whose static public key is $\pk$, we first compute the note public key $(\npk, R)$ of the receiver as described in Section~\ref{sec:protocol-keys}. Next, we need to set the type of the transaction: if the transaction is transparent, we set $\type = 0$, and if the transaction is obfuscated, we set $\type = 1$. We also set $\vnew{}$ to the amount of money of the new note $\Note{}$. Finally, we need to commit to $\vnew{}$, and encrypt the opening as well. To do so, we first set a blinding factor for the commitment and a nonce for the encryption:

\begin{itemize}
	\item If $\type = 0$, set $s = 0$ and $\nonce = 0$.
	\item If $\type = 1$, set $s \leftarrow \F_{t}$ and $\nonce \leftarrow \F_t$.
\end{itemize}

and compute the value commitment $\com = \mathsf{Com}_\ck(v;s)$. Then, we encrypt the opening of $\com$:

\begin{itemize}
	\item If $\type = 0$, then set $\enc = \vnew{}$.
	\item If $\type = 1$, then 
	$\enc = \Enc_{\kdh} (\vnew{}||s; \nonce)$.
\end{itemize}

Now, we can set the new note to \[\Note = \{\type, \com, \nonce, \enc, \npk, R\}.\]

\begin{figure}[h]
	\centering
	\setlength{\fboxsep}{5pt}%
	\setlength{\fboxrule}{0.3pt}%
	\fbox{
		\includegraphics[width=460pt,draft=false]{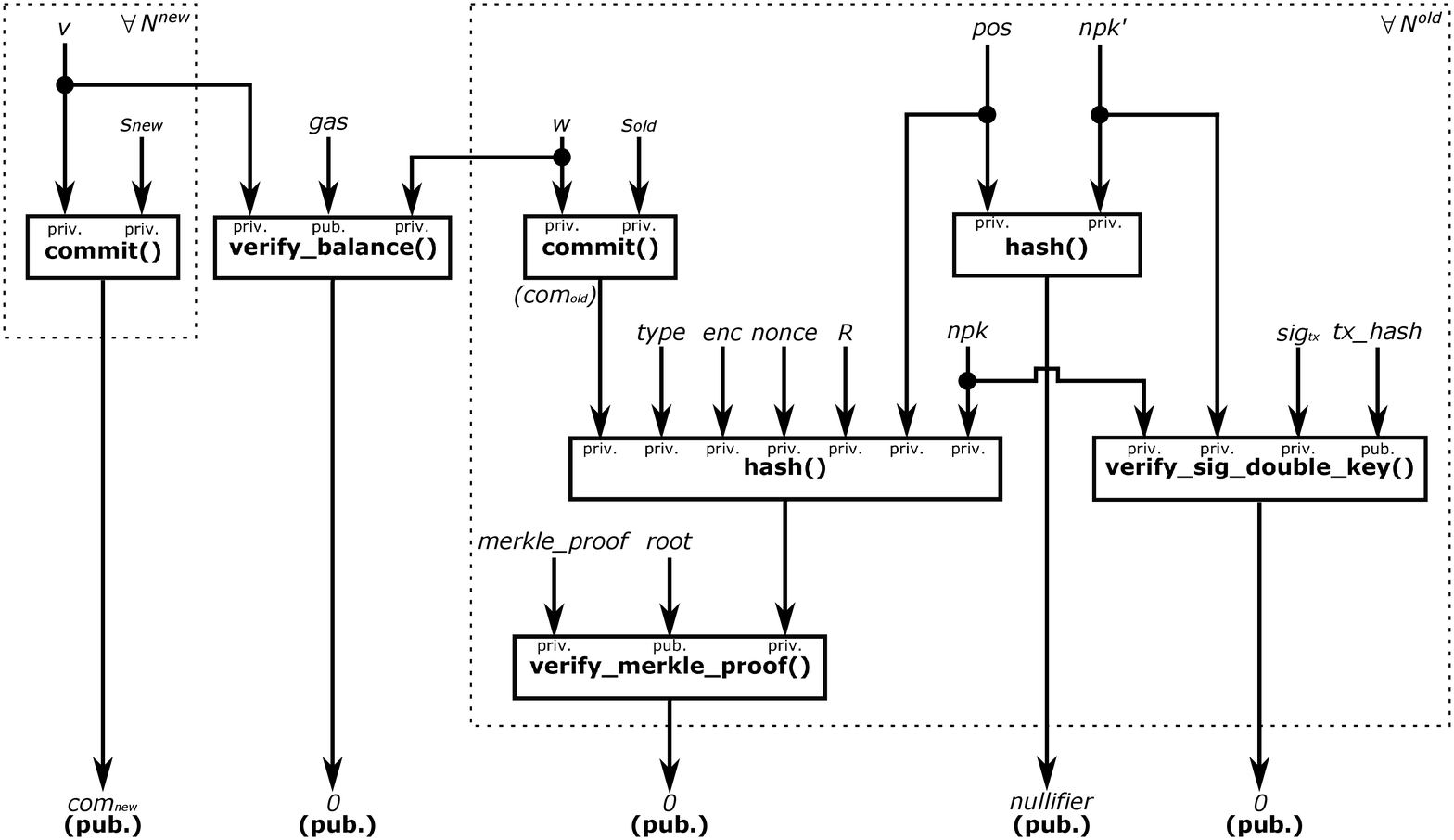}}
	\caption{Arithmetic circuit for a Dusk transaction proof.}
	\label{fig:circuit}
\end{figure}

The next step is to compute a ZKP using the circuit depicted in Figure~\ref{fig:circuit} to prove the following elements:

\begin{itemize}
	\item \textbf{Membership}: the sender must prove that every $\mathsf{N}\in\{\nold{i}\}_{\mathsf{i} = 1}^{\mathsf{m}}$ is included in the Merkle tree of notes. To do so, the sender provides a Merkle proof for $\hp(\mathsf{N})$, and the circuit verifies the Merkle proof in \texttt{verify\_merkle\_proof()}. We observe that all these inputs are private and hence, the proof will not reveal which note is being spent, only that it belongs to the Merkle tree of notes. 

	\item \textbf{Ownership}: the sender must prove that they hold the note secret key $\nsk$ of every note $\mathsf{N}\in\{\nold{i}\}_{\mathsf{i} = 1}^{\mathsf{m}}$. 
	Instead of including their private key as an input to the circuit and computing $\npk$ inside, the sender proves (using the \texttt{verify\_signature()} box inside the circuit) that they can sign a message with that key. In this case, they use the double-key Schnorr signature scheme to sign the hash of the transaction. 

	\item \textbf{Nullification}: the sender must prove that $\nullifier=\hp(\npk' || \pos)$. Note that the sender provides the nullification key $\npk' = \nsk G'$ as an input to the circuit and not the note secret key $\nsk$. As we just explained, the double-key Schnorr signature guarantees that $\npk'$ is indeed $\nsk G'$. The result of the \texttt{hash()} box is the nullifier, which is a public output of the circuit that is later included as part of the transaction.
	
	\item \textbf{Balance integrity}: the \texttt{verify\_balance()} box checks that 
	\begin{equation}\label{eq:balance}
		\sum_{i = 1}^m \vold{i} - \sum_{i = 1}^n \vnew{i} - \mathsf{gas} = 0,
	\end{equation} where $\mathsf{gas}$ is the maximum amount of gas that the sender is willing to pay for the transaction.
\end{itemize}

Observe that using the double-key Schnorr signature as proof that the user holds $\nsk$, allows users to delegate the generation of the ZKP to a partially trusted third party, that is, a \textit{proof helper}. This delegation would require the user to entrust the old and new values of the notes to the proof helper, but not their secret key. 

Finally, the remaining checks not verified inside the circuit are performed by the network. For instance, checking that the nullifier included in the transaction matches the output of the circuit, or that the note has the right typeset.

\section{Our Solution: Citadel}
\label{sec:oursolution}
In this section, we introduce \verb!Citadel!. We start with an overview of the protocol, its elements, and its use cases. Then, we introduce the main element required by our solution: a private NFT model for Dusk. Finally, we detail our novel SSI system, \verb!Citadel!.

\subsection{Overview}

In this section, we describe a novel protocol for authenticating in several services but preserving at the very same time our privacy. Imagine we want to buy a ticket for a concert. Nowadays, what we would do in most cases is to buy the ticket using a web page that will get information about our browser, our credit card, our bank account, ourselves... Moreover, they will charge us a fee for the ticket management service, and probably they will charge a fee to the concert promotor as well. Furthermore, as soon as we show the ticket to the concert, they will be able to link our image to the information previously gathered.

Using \verb!Citadel!, we can do much better. In particular, we want to buy a license (a.k.a. a right) to use some service (e.g. the ticket of the example above is a license, to be used in a concert, which is a service) without leaking any information about us. Furthermore, we want to use such a service as many times as permitted by the SP, without them being able to link our activity, or learn our identity. Moreover, we want a decentralized system that does not rely on third parties to manage our identities and licenses.

To achieve the aforesaid features, we first rely on Dusk Network as the decentralized framework that our solution is based. Then, we need a way to privately share assets among users of the Dusk Network. For this reason, in Section \ref{sec:privatenft} we design a novel and private NFT model for Dusk. Finally, we need a way to anonymously prove ownership of our acquired licenses (i.e. the NFTs), so we introduce our solution in Section \ref{sec:citadel}.

\subsection{A Private NFT Model for Dusk}
\label{sec:privatenft}

As described in Section \ref{sec:buildingblocks}, coins in Dusk are represented as notes, and they can be either transparent note ($\type = 0$), or obfuscated note ($\type = 1$). Here we introduce two new types of notes: transparent NFT note ($\type = 2$) and obfuscated NFT note ($\type = 3$). 

As stated previously, a user willing to spend Dusk notes needs to mint new notes while nullifying the old ones. Like this, a user can spend one note, and create a new note for the receiver, and another one with the change for themselves. When willing to mint a new NFT, a user will need to execute the minting contract where a Dusk note will be used to pay for the contract gas (and thus, it will be nullified), and a new note will be created to receive the change. Additionally, a new NFT note will be created. The creation of an NFT note does not need to be part of the ZKP circuit, as it is not involved in the balance to be verified. As such, it is enough to include the new NFT note in the transaction. A note representing an NFT contains the same data as other notes do, but in this case, what before was the note value in Dusk coins $v$, now is the $\nftpayload$ of the NFT. To mint a new note, we first compute the note public key $\npk$ and the value $R$ of the receiver as described in Section~\ref{sec:protocol-keys}, plus the symmetric key $\kdh$. Then, we set the parameters of each new NFT note $\mathsf{N}$ by executing a function 
\[\mintnft(\npk, R, \nftpayload, \kdh)\]
whose workflow is described in Algorithm \ref{alg:mintnft}.

\begin{algorithm}
\SetAlgoLined
\textbf{Inputs:} \par
$(\npk, R)$: the public note key of the receiver $\npk$ and the related value $R$.\\
$\nftpayload$: a value being the desired content of our NFT note $\Note{}$.\\
$\ksym$: a symmetric encryption key.\break

\textbf{Algorithm:}
	\begin{enumerate}
		\item Set the type of note.
		\begin{itemize}
			\item If the NFT note is transparent, set $\type = 2$.
			\item If the NFT note is obfuscated, set $\type = 3$.
		\end{itemize}
		\item Set a nonce for the encryption.
		\begin{itemize}
			\item If $\type = 2$, set $\nonce = 0$.
			\item If $\type = 3$, set $\nonce \leftarrow \F_t$.
		\end{itemize}
		\item Encrypt the $\nftpayload$.
		\begin{itemize}
			\item If $\type = 2$, then set $\enc = \nftpayload$.
			\item If $\type = 3$, then 
			$\enc = \Enc_{\ksym} (\nftpayload; \nonce)$,
			where $\Enc()$ is a symmetric encryption function. 
		\end{itemize}
		\item Set the new NFT note to \[\Note = \{\type, \enc, \nonce, R, \npk\}.\]
	\end{enumerate}
	\caption{Minting algorithm for private NFTs.}
	\label{alg:mintnft}
\end{algorithm}

As described previously, users willing to spend notes have to nullify them, a process that involves providing a ZKP whose circuit computes the hash of the note. In this process, the parameter $\type$ is set to private, as it is not relevant information for the protocol. \textbf{It is of paramount importance} to notice that after deploying this model, this parameter has to be public. Otherwise, an adversary could spend an NFT note pretending to be spending a regular note and would be able to create huge amounts of money out of the blue.

On the other hand, and as described in this section, the changes in the whole protocol are minimal. As such, deploying our model to the current system should be trivial.

% Furthermore, it is interesting to provide a way to prove ownership of a given NFT. Let us have our prover $\mathcal{P}$ willing to prove ownership of an NFT to a verifying party $\mathcal{V}$. A challenge-response protocol for achieving this purpose is described as follows.

% \begin{protocol}{Proving ownership of an NFT note.}
% \label{pro:citadel}
% \textbf{Environment:} \par
% A prover $\mathcal{P}$ knowing $\nsk$ willing to prove ownership of an NFT with note public key $\npk$. \\
% A verifying party $\mathcal{V}$ that wants to be convinced of $\mathcal{P}$'s statement. \\

% \\
% \textbf{Protocol:}
% 	\begin{enumerate}
% 		\item $\mathcal{V} \rightarrow \mathcal{P}$ : $r \leftarrow \F_t$ \$
% 		\item $\mathcal{P} \rightarrow \mathcal{V}$ : $\sig = \sign_{\nsk}(r)$ 
% 		\item $\mathcal{V} \rightarrow$  : $1/0 \leftarrow \verify_{\npk}(\sig, r)$ 
% 	\end{enumerate}
% \end{protocol}

\subsection{Description of Citadel}
\label{sec:citadel}

Now, we are going the introduce all the details about \verb!Citadel!. Then, we will detail its security analysis, and finally, we will perform some experiments in order to get benchmarks of the protocol.

\subsubsection{Protocol Details}

\begin{figure}[h]
	\centering
		\includegraphics[width=390pt,draft=false]{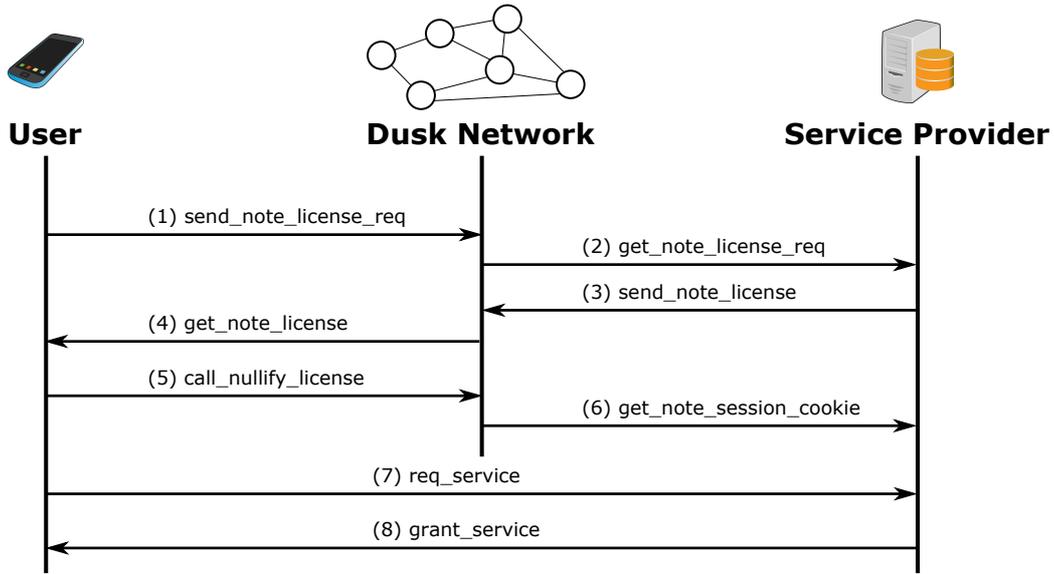}
	\caption{Overview of the protocol messages exchanged between the user, the Dusk Network, and the SP.}
	\label{fig:protocol}
\end{figure}

Let us have a user willing to pay a service provider SP for a license to use their service, and willing to anonymously prove ownership of this license afterward. First, the user will execute a payment in the Dusk Network addressed to the SP, including into the transaction the required information to receive the license. Upon receiving the payment, the SP will send back a license to the user, using the same Blockchain. In order to use the license, the user will have to call a smart contract deployed in the Dusk Network, called the \textit{license contract}. Essentially, the user will provide a proof that demonstrates that they own a valid license, the license contract will verify the proof, and will append a license nullifier to a Merkle tree of nullifiers. By means of a session cookie included in the same contract call, and addressed to the SP, the user will be able to request the service using an off-chain and secure channel. The workflow is depicted in Figure \ref{fig:protocol}, and described with full details in the following protocol.

\begin{figure}[h]
	\centering
	\setlength{\fboxsep}{5pt}%
	\setlength{\fboxrule}{0.3pt}%
	\fbox{
		\includegraphics[width=460pt,draft=false]{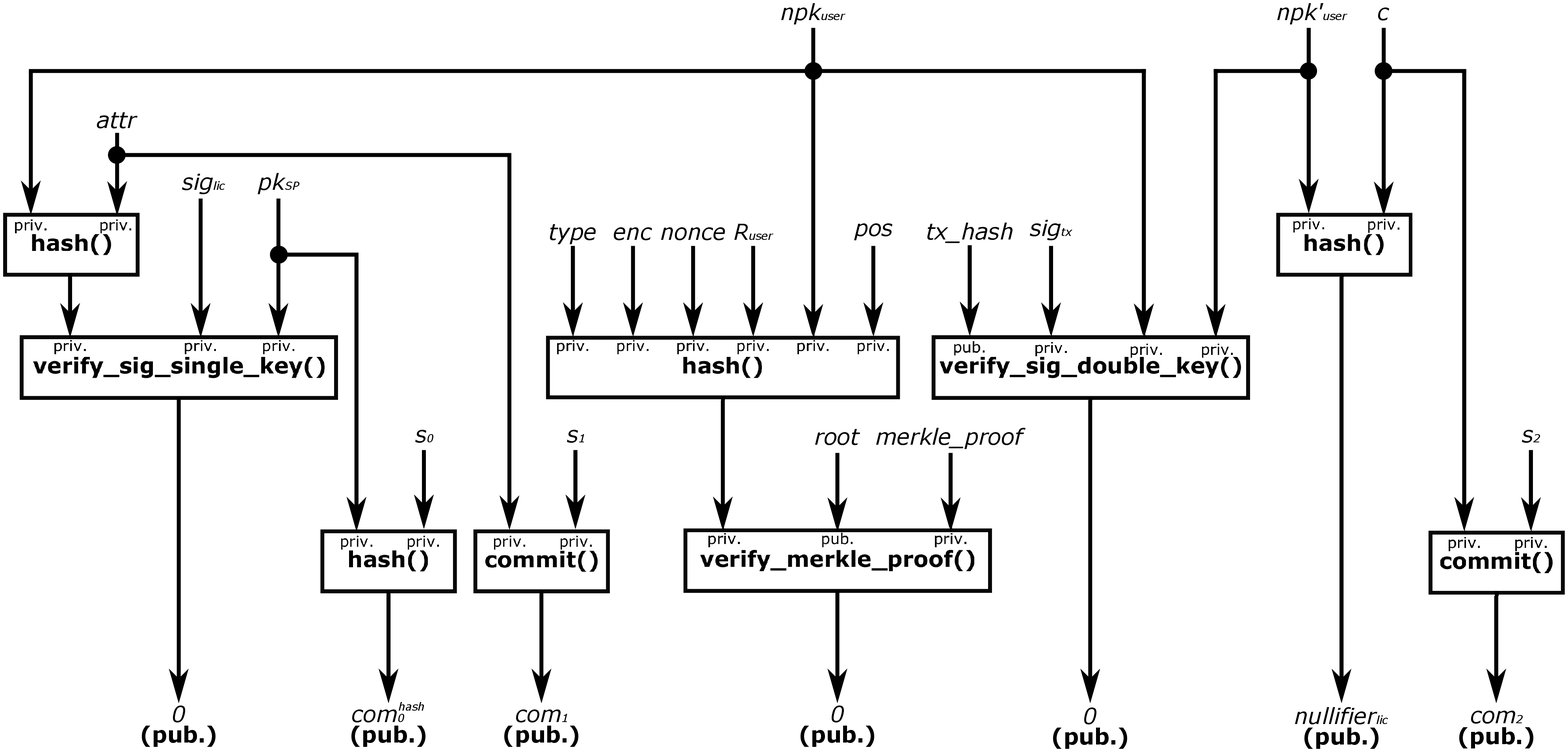}}
	\caption{Arithmetic circuit for proving a license's ownership.}
	\label{fig:circuit_prove_nft}
\end{figure}

\begin{protocol}{Citadel.}
\label{pro:citadel}
\textbf{Environment:} \par
A Service Provider SP offering a service, and publicly sharing its public key $\pk_{\SP}$. \\
A user willing to use a service provided by the SP. \\
The Dusk Network Blockchain.

\\
\textbf{Protocol:}
	\begin{enumerate}
		\item (\textbf{user}) $\mathsf{send\_note\_license\_req}$ : Compute a note public key $(\npk_{\user}, R_{\user})$ belonging to the user, using the user's own public key, and also an additional key $\ksym_{\user} = \hp(\npk_{\user}, \nsk_{\user})$, by computing first the user's $\nsk_{\user}$. Then, send the required amount of Dusk coins to the SP, in order to pay for the service. Into the same transaction, send an NFT to the SP using the function $\mintnft(\npk_{\SP}, R_{\SP}, \nftpayload, \kdh)$, whose arguments are computed as follows:
		\begin{itemize}
			\item $(\npk_{\SP}, R_{\SP})$ is the SP's note public key, computed through his public key $\pk_{\SP}$.
			\item $\nftpayload = (\npk_{\user}, R_{\user}, \ksym_{\user})$.
			\item $\kdh$ is computed using the SP's public key.
		\end{itemize}

		\item (\textbf{SP}) $\mathsf{get\_note\_license\_req}$ : Continuously check the network for incoming license requests. Upon receiving the payment from a user, define a set of attributes $attr$ representing the license, and compute a digital signature as follows:

		$$\lsig= \sign_{\sk_{\SP}}(\npk_{\user}, \attr)$$

		\item (\textbf{SP}) $\mathsf{send\_note\_license}$ : Set the $\nftpayload = \{\lsig, \attr\}$, and send the license to the user using the function $\mintnft(\npk_{\user}, R_{\user}, \nftpayload, \ksym_{\user})$.

		\item (\textbf{user}) $\mathsf{get\_note\_license}$ : Receive the note containing the license. 

		\item (\textbf{user}) $\mathsf{call\_nullify\_license}$ : When desiring to use the license, nullify it by executing a call to the license contract. The following steps are performed:

		\begin{itemize}
			\item The user sets a session cookie $\stoken = (\mathsf{s_0}, \mathsf{s_1}, \mathsf{s_2}) \leftarrow \F_t$.
			\item The user creates a new NFT note where $\nftpayload = \stoken$, and the SP is the receiver.
			\item The user issues the transaction that includes the NFT described in the previous step, by calling the license contract. In this case, the \textsf{tx\_proof} is computed as done in the standard Phoenix model, but into the same circuit, the circuit depicted in Figure \ref{fig:circuit_prove_nft} is appended.
			\item The network validators will execute the smart contract, which verifies the proof. Upon success, the NFT note will be forwarded, and the license nullifier $\lnullifier$ will be added to the Merkle tree of nullifiers.
		\end{itemize}

		\item (\textbf{SP}) $\mathsf{get\_note\_session\_cookie}$ : Receive a note containing the session cookie $\stoken$.

		\item (\textbf{user}) $\mathsf{req\_service}$ : Request the service to the SP, establishing communication using a secure channel, and providing the tuple $(\mathsf{tx\_hash}, \pk_{\SP}, \attr, c, \stoken)$.

		\item (\textbf{SP}) $\mathsf{grant\_service}$ : Grant or deny the service upon verification of the following steps:

		\begin{itemize}
			\item Check whether or not the values $(\attr, \pk_{\SP}, c)$ are correct.
			\item Check whether or not the openings $((\pk_{\SP}, \mathsf{s_0}), (\attr, \mathsf{s_1}), (c, \mathsf{s_2}))$ match the commitments $\com_0^{hash}, \com_1, \com_2$ found in the transaction $\mathsf{tx\_hash}$.
		\end{itemize}

	\end{enumerate}
\end{protocol}

As can be seen, the fact that only the user knows the $\nsk$ required to compute $\lsig$ allows them to prove ownership of the license, by means of the double key Schnorr signature, and this license is verified by proving knowledge of a valid signature verified with the public key of the SP. 

Moreover, we can appreciate in the circuit how the license is linked to the $\npk_{\user}$, and the user also verifies a Merkle proof that proves membership of this note in the Dusk Network. This brings the revocation feature: if under some circumstances the SP no longer accepts some previously issued licenses, they can prove to the network that a given note contains a license issued by them, and under a consensus agreement, it will be removed. As such, after removal, the user will not be able to provide valid proofs for this license anymore. Plus, it can happen that a user receives a license, but the transaction is finally not accepted in the Blockchain (e.g. the transaction proof was not correct, some regulation checks have failed, etc.). In this scenario, the received license will not be valid, and the user indeed will not be able to provide a correct Merkle proof.

Furthermore, the SP might request the user to nullify the license they are using (i.e. this is a single-use license, like entering a concert). This is done through the computation of $\lnullifier$. The deployment of this part of the circuit has two different possibilities:
	\begin{itemize}
		\item If we set $c = 0$ (or directly remove this input from the circuit), the license will be able to be used only once.
		\item If the SP requests the user to set a custom value for $c$ (e.g. the date of an event), the license will be able to be reused only under certain conditions.
	\end{itemize}

It is also interesting to notice that the whole protocol has been designed with perfect integration into Dusk Network. As can be noticed, all the information needed to prove ownership of a license is stored on-chain. As such, a user setting up a new instance of the Dusk wallet will be able to retrieve all the licenses by simply knowing his static secret key, as would be done with the whole Dusk protocol. The same happens with the delegation of the received notes check, where the user delegates the process of checking which notes are addressed to them. In \verb!Citadel!, the user can securely delegate the check of received licenses. Furthermore, proof delegation is also possible, as the user knowing $\nsk$ will use this key to sign a specific transaction, and this cannot be modified.

\subsubsection{Security Analysis}

We start the security analysis of \verb!Citadel! by elaborating on the ZKP scheme to use. The need for a trusted setup is one of the main drawbacks of some ZKP constructions like PlonK, especially when used in scenarios like cryptocurrencies. An untrusty setup where an adversary gets the seed used to compute it would allow them to create false transactions, and this would lead to huge losses of money. In \verb!Citadel!, an untrusted setup would lead to user impersonation, and being able to use others' licenses.

On the other hand, the soundness property of each scheme relies on different security assumptions \cite{assumptions} (e.g. the specific zk-SNARK described in \cite{cryptoeprint:2016:260} uses a strong assumption, the q-Power Knowledge of Exponent ($q$-PKE) assumption). In other words, the security of these schemes relies on the security of elliptic curves, where breaking the security of the selected curve would lead to being able to generate false proofs. In our scenario, the BLS12-381 \cite{zcash} is used. It is estimated to have around 128 bits of security, which complies with the security standards.

We now put the spotlight on the security of the circuit we have designed, which grants the following features:

\begin{itemize}
	\item \textbf{Proof of Ownership:} the circuit used in \verb!Citadel! verifies a signature $\lsig$ of an input message $(\npk_{\user}, \attr)$, using the public key of SP, $\pk_{\SP}$. Also, a double key signature $\tsig$ of a transaction hash $\mathsf{tx\_hash}$ is verified in-circuit, referring to the transaction where the ZKP will be appended.

	$\lsig$ verification ensures that the license attributes are correct, and $\tsig$ ensures that the user owns such a license, as only they can compute $\npkn{\user}$ using the note secret key and compute such a signature, while keeping all these values private, so SP cannot learn the identity of the user. An adversary would not be able to prove ownership as long as $\nsk_{\user}$ is not leaked to them. This is true under the discrete logarithm assumption.
 
	\item \textbf{Proof of Validity:} the fact that $\npk_{\user}$ is part of the signature $\lsig$, ensures that the license is assigned to a specific note of the Dusk Network, and thus, a specific user of this blockchain. The circuit verifies a Merkle proof of the NFT note containing the license, which is included in the Merkle tree of notes. This ensures that the license the user is proving ownership of has been transacted in the Dusk Network, and is a valid license at the moment of issuing the transaction. 

 	An adversary willing to successfully prove ownership of a transferred license would have to craft a new pair $(\npk_{\user}, $attr$)$ that verifies $\lsig$. This is infeasible under the discrete logarithm assumption. Furthermore, the crafted $\npk_{\user}$ would have to be a collision verifying the Merkle proof.

 \item \textbf{Unlinkability:} the user sends the one-time key pair $(\npk_{\user}, R_{\user})$ to the SP, instead of the public key $\pk$. The fact that the information about the user learned by the SP is a set of one-time values ensures that the identity of the user sending these values cannot be linked to other activities done in the network. The key $\npk_{\user}$ is computed from the value $\nsk_{\user}$, which is kept secret and used only one time. As there are no other values involved in the process that identifies the user, they cannot be linked to the user's identity. This is true as long as the user does not reuse $\nsk_{\user}$. On the other hand, $\npk_{\user} = \hb(rA)G + B$, where $r$ is sampled at random and $(A, B)$ is the user's public key. As both $\hb(rA)G$ and $B$ are only known by the user, there is no way an adversary can learn $B$, because $\npk_{\user}$ can be decomposed in many ways.

 From the point of view of the network, there is unlinkability as well: when issuing the transaction, no one is able to link the nullified license to the SP, as the $\pk_{\SP}$ is blinded by committing to this value using the \textsf{hash()} function and a random value $\mathsf{s_0}$. An adversary would not be able to learn $\pk_{\SP}$ as long as the randomness involved in the hashing process is not leaked to them. This is true assuming that the hashing function is collision-resistant. On the other hand, both $\attr$ and $c$ could leak information about the service and the user. For this reason, we commit to these values (as they are scalars instead of points, we can use the Pedersen Commitment which requires fewer constraints than the hash function). An adversary would not be able to learn $(\attr, c)$ as long as the random values involved in the commitments are not leaked to them. This is true under the discrete logarithm assumption, which holds for the Pedersen commitment.
 
 \item \textbf{Decentralized Nullification:} the circuit computes the hash of $\npkn{\user}$ and a public challenge $c$, resulting in $\lnullifier$. The format of the $c$ value could change in different scenarios. Taking the example of proving ownership of a ticket for an event, ideally, $c$ would be the date of such an event. If a function checking if a given nullifier has been previously seen, results in the equation $\isseen(\lnullifier, \pnullifiers[]) \stackrel{?}{=} 1$ holding, it means that someone already entered the event with the same license. As such, we ensure that a user cannot use the same license multiple times, nor compute valid proofs for other users. $\npkn{\user}$ is fixed in advance, as such, $\lnullifier$ will always be the same for a given public input $c$, which needs to be validated by the SP.
 
\item \textbf{Attribute Blinding:} As described previously, the user provides an opening for the commitment $\com_1$ to the SP, thus leaking the $\attr$ value. An adversary would not be able to provide a valid opening as long as the randomness involved in the commitment of $\attr$ is not leaked to them. This is true under the discrete logarithm assumption, which holds for the Pedersen commitment.

Depending on the use case, it could be desirable that the values involved in $\attr$ are kept totally or partially private. In this scenario, and as suggested in the \verb!FORT! protocol, the user could instead provide an additional proof of knowledge, proving to the SP that they know the opening of $\com_1$. As an example, a Bulletproof is a kind of ZKP allowing to prove knowledge of a value that lies within a certain range. 
 
\end{itemize}

\subsubsection{Benchmarks}

We use \verb!dusk-plonk! to code the circuit used in our solution. Like this, we get the number of constraints of its different elements and thus, the efficiency of computing (and verifying) proofs. Our circuit needs four main functions:

\begin{itemize}
 \item \textsf{hash():} we use the Poseidon hash function, which uses \textbf{977 contraints} when hashing 1 input. The amount of constraints increases depending on the number of inputs.

 \item \textsf{commit():} the Pedersen commitment requires \textbf{527 constraints}.

 \item \textsf{verify\_sig\_single\_key():} Dusk uses the Schnorr proof signature scheme over BLS12-381, which uses \textbf{3388 constraints} when using the single key version.

 \item \textsf{verify\_sig\_double\_key():} Dusk uses the Schnorr proof signature scheme over BLS12-381, which uses \textbf{6645 constraints} when using the double key version.

 \item \textsf{verify\_merkle\_proof():} the circuit needs to verify a Merkle proof. Dusk uses Merkle trees of depth 17, as such, our solution will need to compute 17 Poseidon hashes. This sums up to \textbf{17807 constraints}. 
\end{itemize}

We implemented our circuit using a total amount of \textbf{34861 constraints}. Here, we need to add the constraints needed to compute the default \textsf{tx\_proof}, which are \textbf{31486 constraints} for nullifying one note. We benchmarked both the prover and verifier using an Apple Silicon M1 CPU. The prover takes \textbf{16.232 seconds} to compute the proof, and the verifier \textbf{0.007 seconds} to verify it.

Plus, it has to be taken into account that an advantage of \verb!Citadel! is that licenses can be nullified much before being used. This means that long ZKP proving times will not have a big impact on the performance of the protocol even when using devices with low computing resources. Nonetheless, and as mentioned previously, computations can be delegated as done in the standard Phoenix model, so high-performing CPUs will compute the proofs even in less time.

\section{Conclusions and Future Work}
\label{sec:conclusions}
In this paper, we have introduced two main contributions. We have described a private-by-design NFT model for the Dusk Network Blockchain. This allows issuing of unique assets in the network, that remain private to anyone but the receiver. We also showed how this approach perfectly integrates into the current transaction model of Dusk. 

Then, we introduced \verb!Citadel!, which uses the NFT model to issue licenses to users, whose ownership can be privately demonstrated afterward. Among all the features described in the protocol details, we demonstrated how the main problems in previous solutions are solved: licenses cannot be traced on-chain, and their usage can be nullified in a decentralized manner by means of the Blockchain itself. As such, our solution serves the privacy concerns of the users, while the SPs have the guarantee that the licenses they issue will not be misused (e.g. used more times than what is permitted). After detailing the security behind our protocol, we provided a proof-of-concept implementation of the proving system, along with some benchmarks and deployment specifications that prove its feasibility in a real scenario.

Beyond all the use cases and examples described in this work, our solution can also be seen as a framework to be adapted and modified according to the needs that specific applications can have. Working in more specific applications and being able to deploy \verb!Citadel! in real scenarios would be interesting future work.

\phantomsection
\addcontentsline{toc}{section}{Acknowledgements}
\section*{Acknowledgements}
This work has been funded by Dusk Network. We also want to thank Marta Bellés and Javier Silva from Dusk Network for their insightful comments on this work.

\phantomsection
\addcontentsline{toc}{section}{References}
\bibliographystyle{unsrt}
\bibliography{bibtex}

\end{document}